\documentclass[aps, preprint, showpacs,showkeys,superscriptaddress]{revtex4-1}

\usepackage{graphicx}
\usepackage{epsfig}  
\usepackage{epsf}    
\usepackage{dcolumn}
\usepackage{bm}
\usepackage{dcolumn}
\usepackage{textcomp}
\usepackage[tbtags]{amsmath}
\usepackage{amsfonts}
\usepackage{float}
\usepackage{subfig}
\usepackage[]{hyperref}
  \hypersetup{
  unicode=false,          
  pdftoolbar=true,        
  pdfmenubar=true,        
  pdffitwindow=true,     
  pdfstartview={FitH},    
  pdfsubject={Getting the best out of T2K and NOvA},   
  pdfnewwindow=true,      
  pdfcreator={RevTeX},
  colorlinks=true,       
  linkcolor=red,          
  citecolor=blue,        
  urlcolor=blue,           
  }
\usepackage{hypcap}
\usepackage{appendix}

\newcommand{\beq}{\begin{equation}}
\newcommand{\eeq}{\end{equation}}
\newcommand{\beqa}{\begin{eqnarray}}
\newcommand{\eeqa}{\end{eqnarray}}

\newcommand{\tx}{{\theta_{12}}}
\newcommand{\ty}{{\theta_{13}}}
\newcommand{\tz}{{\theta_{23}}}

\newcommand{\dl}{{\Delta_{31}}}
\newcommand{\ds}{{\Delta_{21}}}

\newcommand{\dcp}{\delta_{\mathrm{CP}}}
\newcommand{\nova}{NO$\nu$A~}

\newcommand{\dchsq}{\Delta\chi^2}

\begin{document}

\preprint{APS/123-QED}

\title{Non-unitary neutrino mixing in the NO$\nu$A near detector data}

\author{Ushak Rahaman}
\email{ushakr@uj.ac.za}
\author{Soebur Razzaque}%
 \email{srazzaque@uj.ac.za}
\affiliation{Centre for Astro-Particle Physics (CAPP)\\ and Department of Physics,\\ University of Johannesburg, PO Box 524, Auckland Park 2006,\\ South Africa}%

\date{\today}

\begin{abstract}
The $\nu_\mu \to \nu_e$ oscillation probability over short baseline ($\lesssim 1$~km) would be negligible in case the mixing matrix for three active neutrinos is unitary. 
However, in case of non-unitary 
mixing of three neutrinos, this probability would be non-negligible due to the so-called ``zero distance'' effect.
Hence, the near detector of the accelerator experiments such as \nova can provide 
strong constraints on the parameters of the non-unitary mixing with very large statistics.
By analyzing the \nova near detector data we find that the non-unitary mixing does not improve fits to the $\nu_e$ or $\nu_\mu$ events over the standard unitary mixing. This leads to constraints on the non-unitary parameters: $\alpha_{00}>0.911$, $|\alpha_{10}|<0.020$ and  $\alpha_{11}>0.952$ at 90\% C.L.
A combined analysis with the near and far detector data does not change these constraints significantly.
\end{abstract}

\maketitle

\section{Introduction}
The latest results from the \nova \cite{Himmel:2020, NOvA:2021nfi} and T2K \cite{Dunne:2020} show that there exists a tension between the two experiments. Their best-fit points in the $\sin^2\tz-\dcp$ plane, with unitary mixing hypothesis, are far apart from each other and without any overlap between their respective $1\, \sigma$ confidence regions.
Moreover, while separate analyses of the \nova and T2K data prefers normal hierarchy (NH), the combined analysis prefers inverted hierarchy (IH) over NH \cite{Kelly:2020fkv}. Several papers have tried to resolve this tension with the help of beyond the standard model (BSM) physics. Ref.~\cite{Chatterjee:2020kkm, Denton:2020uda} 
considered neutral current non-standard interaction during propagation, while Ref.~\cite{Rahaman:2021leu} 
considered CP-violating Lorentz invariance violation during propagation. However, the new physics signatures are not stronger than $1\, \sigma$ significance in those analyses and no other neutrino oscillation experiments have observed those signatures yet.  Ref.~\cite{Miranda:2019ynh} instead considered non-unitary $3\times 3$ mixing as a solution to the tension between the \nova and T2K data. Non-unitary mixing implies presence of extra neutrino generations, which is in accordance with the LSND~\cite{Aguilar:2001ty} and MiniBooNe~\cite{Aguilar-Arevalo:2018gpe} results. However, a recent analysis~\cite{Forero:2021azc} shows that data from the short baseline reactor neutrino experiments 
strongly disfavors non-unitary mixing. 
Therefore, further analyses of more available data are needed to search for signatures of non-unitary mixing  
in long baseline neutrino oscillation experiments.
Here, we analyze the \nova near detector (ND) data with non-unitary neutrino mixing hypothesis for the first time.

The accelerator neutrino energy is in the  $1-10$ GeV range. Since the neutrino oscillation probability depends on $\dl L/E$, where $\dl = m_3^2-m_1^2$ is the mass-squared-difference between the neutrinos of masses $m_3$ and $m_1$, $L$ is the length of the baseline and $E$ is the neutrino energy. The oscillation probability for the standard unitary mixing, driven by $\dl$ at this energy range, is negligible at $L\sim 1$ km. This is not true, however, in the case of non-unitary mixing. The non-unitary $3\times 3$ mixing matrix is defined as \cite{Escrihuela:2015wra}
\begin{equation}
N=N_{\rm NP}U_{\rm PMNS}= \left[ {\begin{array}{ccc}
   \alpha_{00} & 0 & 0 \\
   \alpha_{10} & \alpha_{11} & 0 \\
   \alpha_{20} & \alpha_{21} & \alpha_{22}
  \end{array} } \right] U_{\rm PMNS} \,,
\end{equation}
where $U_{\rm PMNS}$ is the unitary PMNS mixing matrix. In the $N_{\rm NP}$ matrix, the diagonal term(s) must deviate from unity and/or the off-diagonal term(s) deviate from zero to allow for the non-unitarity effect. The $\nu_\mu \to \nu_e$ oscillation probability at $L=0$ in this case is given by \cite{Escrihuela:2015wra}
\begin{equation}
P_{\mu e}^{\rm NU} (L=0)=\alpha_{00}|\alpha_{10}|^2
\label{prob-nu-0}
\end{equation}
Therefore, in the case of non-unitary mixing, it is possible to observe neutrino oscillation even over a very short distance, which is called the ``zero distance'' effect. This effect can be tested with data from the short baseline neutrino oscillation experiments as well as the near detector of the long baseline experiments like \nova and T2K.
Although the analysis in Ref.~\cite{Forero:2021azc} shows that the short baseline neutrino experiments like NOMAD \cite{NOMAD:2003mqg}, NuTeV \cite{NuTeV:2002daf} strongly disfavors non-unitary mixing, the reactor neutrino fluxes have large uncertainties~\cite{Dentler:2017tkw}. Accelerator neutrino fluxes are better known. In this paper, we have looked for the signal of non-unitary mixing in the \nova ND data. Besides, we do a combined analysis of the \nova ND and far detector (FD) data to test the two mixing schemes, which is unique for the same experiment. A similar analysis with the T2K near and far detectors would be equally interesting, however, T2K ND data is not publicly available~\cite{Dunne-com}.


The \nova ND is a 290 ton totally active scintillator detector, placed at 100~m underground (to minimize background fluxes) and 1.9~mrad off-axis at approximately 1 km away from the source at the Fermilab~\cite{NOvA:2018gge}. In the case of unitary mixing, the ND will not observe any $\nu_\mu \to \nu_e$ oscillation at such a short distance and the observed electron events will come from the contamination in the $\nu_\mu$ beam.
However, as explained in equation~(\ref{prob-nu-0}), there will be an excess of electron events due to the $\nu_\mu \to \nu_e$ conversion in case of non-unitary mixing.
Effects of non-unitary mixing at the ND will be sensitive to the parameters $\alpha_{00}$ and $|\alpha_{10}|$, and also on $\alpha_{11}$ because of the condition  $|\alpha_{10}|\leq \sqrt{(1-\alpha_{00}^2)(1-\alpha_{11}^{2})}$  \cite{Antusch_2014, Escrihuela:2016ube}. There is essentially no dependence on the phase $\phi_{10}$ associated with $|\alpha_{10}|$ in the ND data. 
For our analyses we have used $1.10 \times 10^{21}$ ($1.18 \times 10^{21}$) protons on target (POTs) data in the neutrino (antineutrino) mode for the ND, and $1.36 \times 10^{21}$ ($12.5 \times 10^{20}$) POTs data for the FD~\cite{Himmel:2020, NOvA:2021nfi}. 

In this paper, we have discussed the analysis in Section~\ref{analysis}, and presented the result in Section~\ref{result}. The conclusion has been drawn in Section~\ref{conclusion}. We show the oscillation probabilities and fits to the event distributions in the Appendix. 

\section{Analysis of \nova data}
\label{analysis}
We calculate the theoretical muon and electron event rates and the $\chi^2$ between the theoretical and experimental data by using the software GLoBES \cite{Huber:2004ka, Huber:2007ji}. We have modified the probability code of GLoBES so that it can handle non-unitary mixing for simulating theoretical events. A detailed algorithm used for the probability calculation with non-unitary mixing has been given in Ref.~\cite{Miranda:2019ynh}. We have also fixed the efficiencies of the electron and muon events for each energy bin according to the expected event rates at the ND and FD which are provided by the \nova Collaboration~\cite{NOvA:2021nfi}. For the $\chi^2$ analyses, we have kept $\sin^2\tx$ and $\ds$, where $\ds = m_2^2 - m_1^2$, at their best-fit values of $0.304$ and $7.42\times 10^{-5}\, {\rm eV}^2$, respectively~\cite{Esteban:2018azc}. We have varied $\sin^2 \ty$ and $\sin^2 \tz$ in their respective $3\, \sigma$ ranges given in  Ref.~\cite{Esteban:2018azc}. We have also varied $|\Delta_{\mu \mu}|$ in its $3\, \sigma$ range around the MINOS best-fit value $2.32\times 10^{-3}\, {\rm eV}^2$ with $3\%$ uncertainty \cite{Nichol:2012}, where $\Delta_{\mu \mu}$ is related to $\dl$ by the following relation \cite{Nunokawa:2005nx}
\begin{eqnarray}
\Delta_{\mu \mu}&=& \sin^2 \tz \dl + \cos^2 \tx \Delta_{32}\nonumber \\
&&+\cos \dcp \sin 2\tx \sin \ty \tan \tx \ds.
\end{eqnarray}
We have varied the CP-violating phase $\dcp$ in its complete range $[-180^\circ:180^\circ]$. For the non-unitary parameters, we have varied $\alpha_{00}$ and $\alpha_{11}$ in the range $[0.7:1]$ and $|\alpha_{10}|$ in the range $[0:0.2]$. We have varied the phase $\phi_{10}$, where $\alpha_{10}=|\alpha_{10}|e^{i\phi_{10}}$, in its complete range $[-180^\circ:180^\circ]$. We have also implemented the condition $|\alpha_{10}|\leq \sqrt{(1-\alpha_{00}^2)(1-\alpha_{11}^{2})}$ throughout the analysis. We have kept the other non-unitary parameters fixed at their unitary values as their effects are negligible.

We have implemented automatic bin-based energy smearing for generated theoretical events as described in the GLoBES manual \cite{Huber:2004ka, Huber:2007ji}.
For this purpose, we have used a Gaussian smearing function for the true neutrino energy $E$
\begin{equation}
R^c (E,E^\prime)=\frac{1}{\sqrt{2\pi}}e^{-\frac{(E-E^\prime)^2}{2\sigma^2(E)}},
\end{equation}
where $E^\prime$ is the reconstructed energy. The energy resolution function is given by 
\begin{equation}
\sigma(E)=\alpha E+\beta \sqrt{E}+\gamma.
\end{equation}
For the \nova FD, we have used $\alpha=0.09\, (0.08)$ and $\beta=\gamma=0$ for $\nu_\mu$ ($\bar{\nu}_\mu$) charge current (CC) events. For $\nu_e$ ($\bar{\nu}_e$) CC events, values for the FD are $\alpha=0.11\, (0.09)$, $\beta=\gamma=0$. For the ND, we have used $\alpha=0.118$, $\beta=\gamma=0$, for both the muon and electron events as well as for both neutrinos and anti-neutrinos. For the FD we have used $8.5\%$ normalization and $5\%$ energy calibration systematic uncertainties for both the $e$-like and $\mu$-like events \cite{NOvA:2018gge}. We calculated the systematic uncertainties at the ND in the following method. We assumed the normalization and energy calibration systematic uncertainties at ND to be $0.085+s_1$ and $0.05+s_2$ respectively for both the $e$-like and the $\mu$-like events. Then we analyzed both ND and FD data together with the standard 3 flavour oscillation hypothesis while both $s_1$, and $s_2$ being varied in the range $[0:0.3]$. For this purpose, we kept the standard oscillation parameters $\sin^2\ty$, $\sin^2\tz$, $\dcp$, and $\dl$ fixed at their best-fit values for NH taken from the analysis of \nova collaboration \cite{Himmel:2020, NOvA:2021nfi}. $\sin^2\tx$, and $\ds$ were fixed at their best-fit values taken from Ref.~\cite{nufit}. We calculated the $\chi^2$ between theory and experiment and found out the minimum $\chi^2$ occurs at $s_1=s_2=0.30$. Thus, we fixed the normalization and energy calibration uncertainties at $38.5\%$ and $30\%$, respectively. 

We have calculated $\chi^2$ for both the NH and IH of the neutrino masses. The minimum $\chi^2$ obtained by varying parameters is subtracted from other $\chi^2$ values to calculate $\dchsq$. During the $\chi^2$ calculation, priors have been added on $\sin^2 \ty$. Further details of the $\chi^2$ analysis are given in Ref.~\cite{Miranda:2019ynh}. Before proceeding to analyze data with non-unitary mixing, we have analyzed data with the standard mixing hypothesis and found that the \nova ND data does not have any sensitivity to the unitary oscillation parameters. This complies with the physics of standard oscillation. In Ref.~\cite{Miranda:2019ynh, Rahaman:2021leu} we presented results from analyzing the FD data with the standard unitary mixing. We matched results of the standard analyses published by the \nova Collaboration to validate our analyses. We present results of our analyses with non-unitary mixing next.

\section{Results and discussion} 
\label{result}
\begin{figure}[h]
\centering
\includegraphics[width=0.45\textwidth]{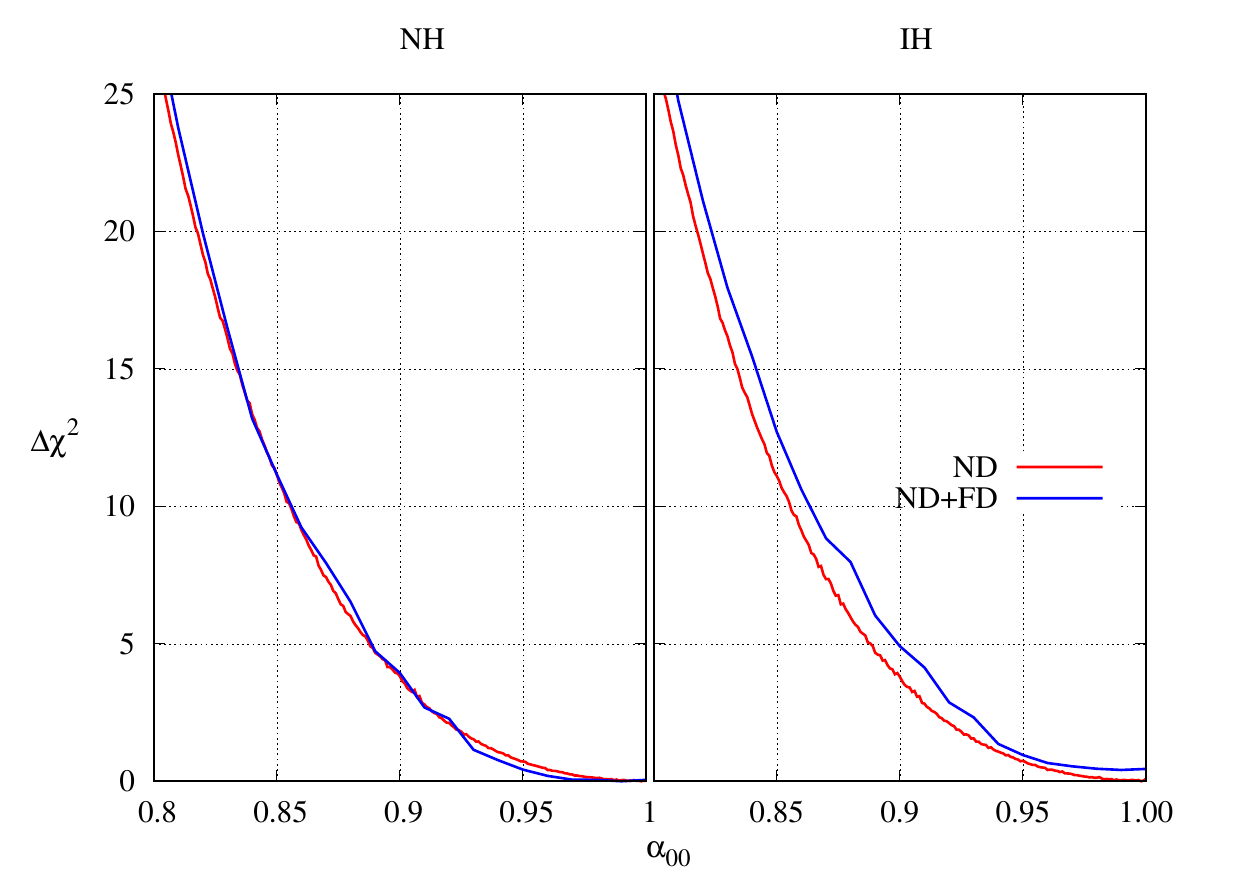}
\includegraphics[width=0.45\textwidth]{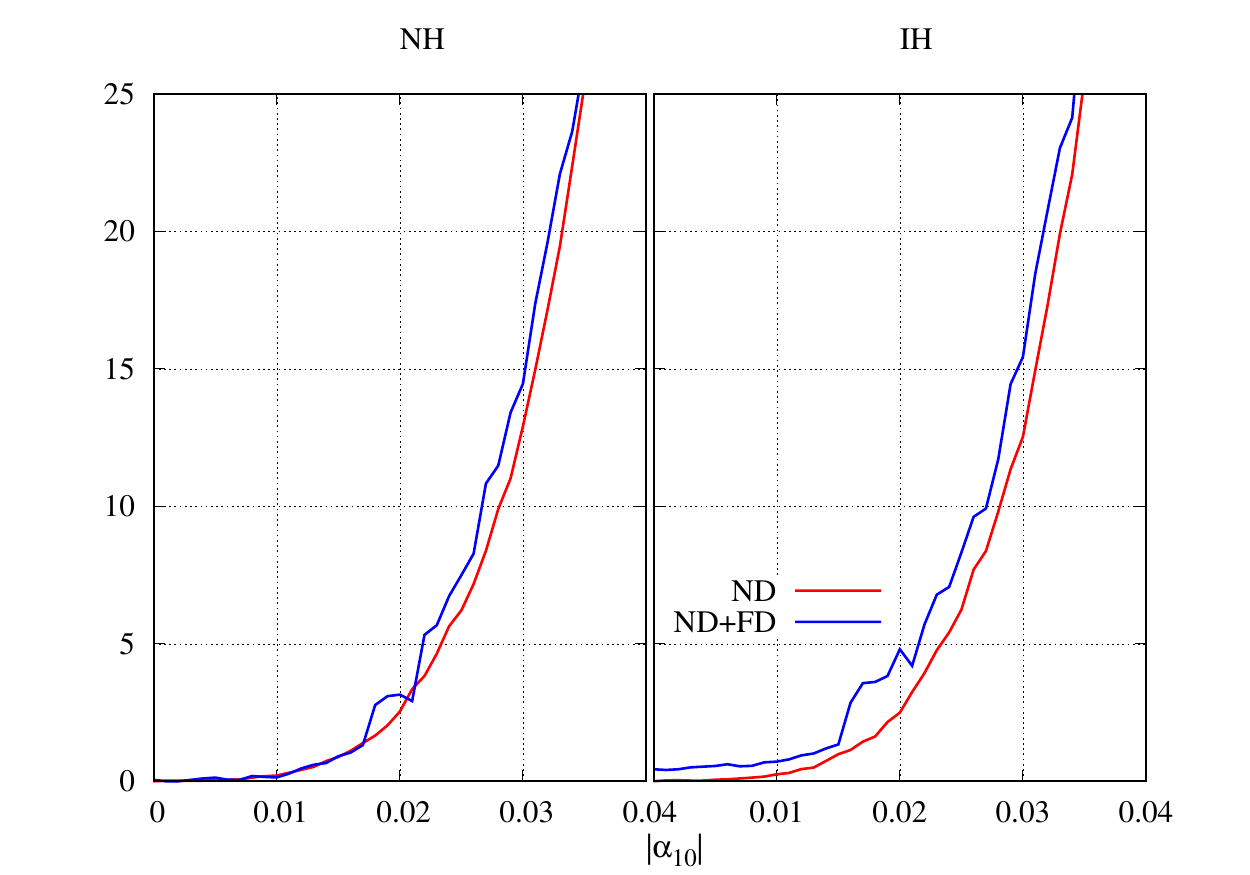}
\includegraphics[width=0.45\textwidth]{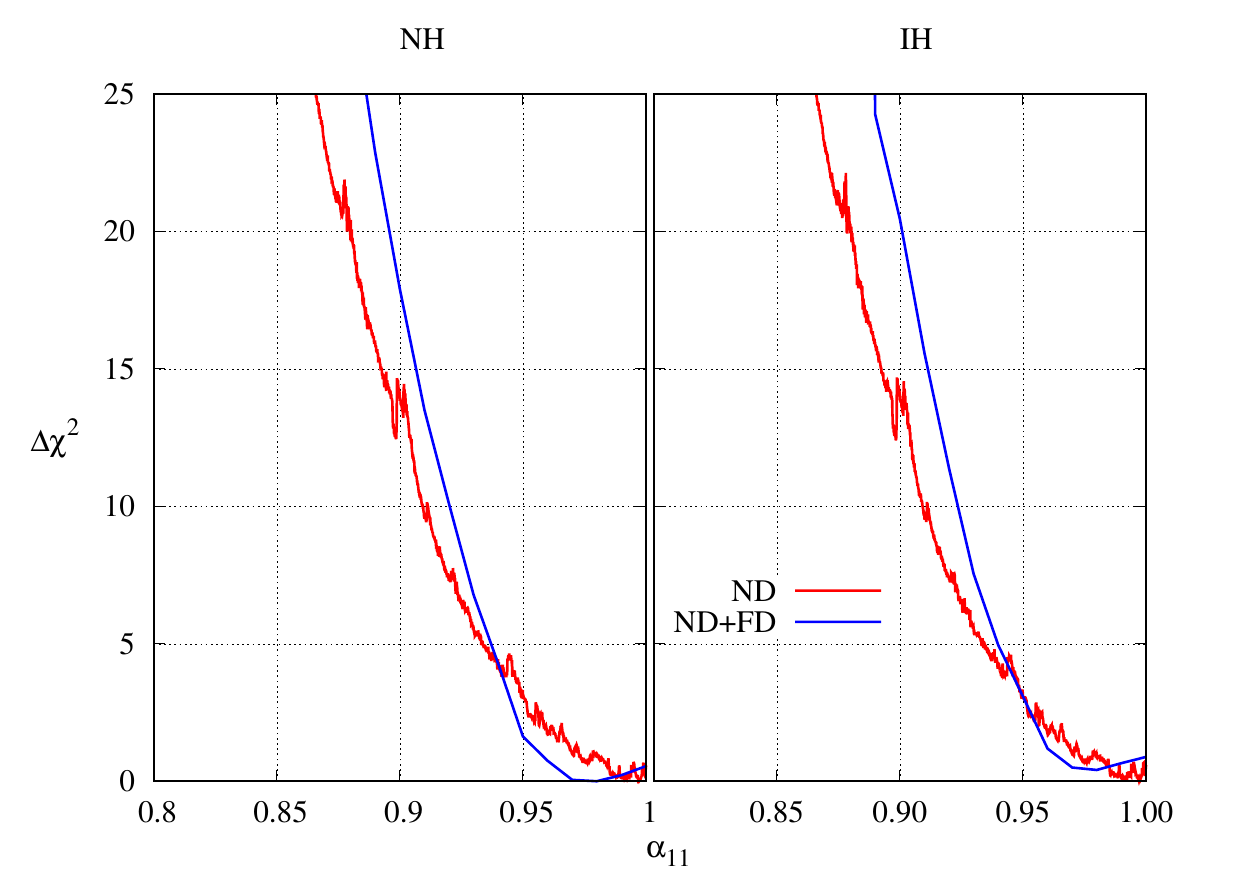}
\includegraphics[width=0.45\textwidth]{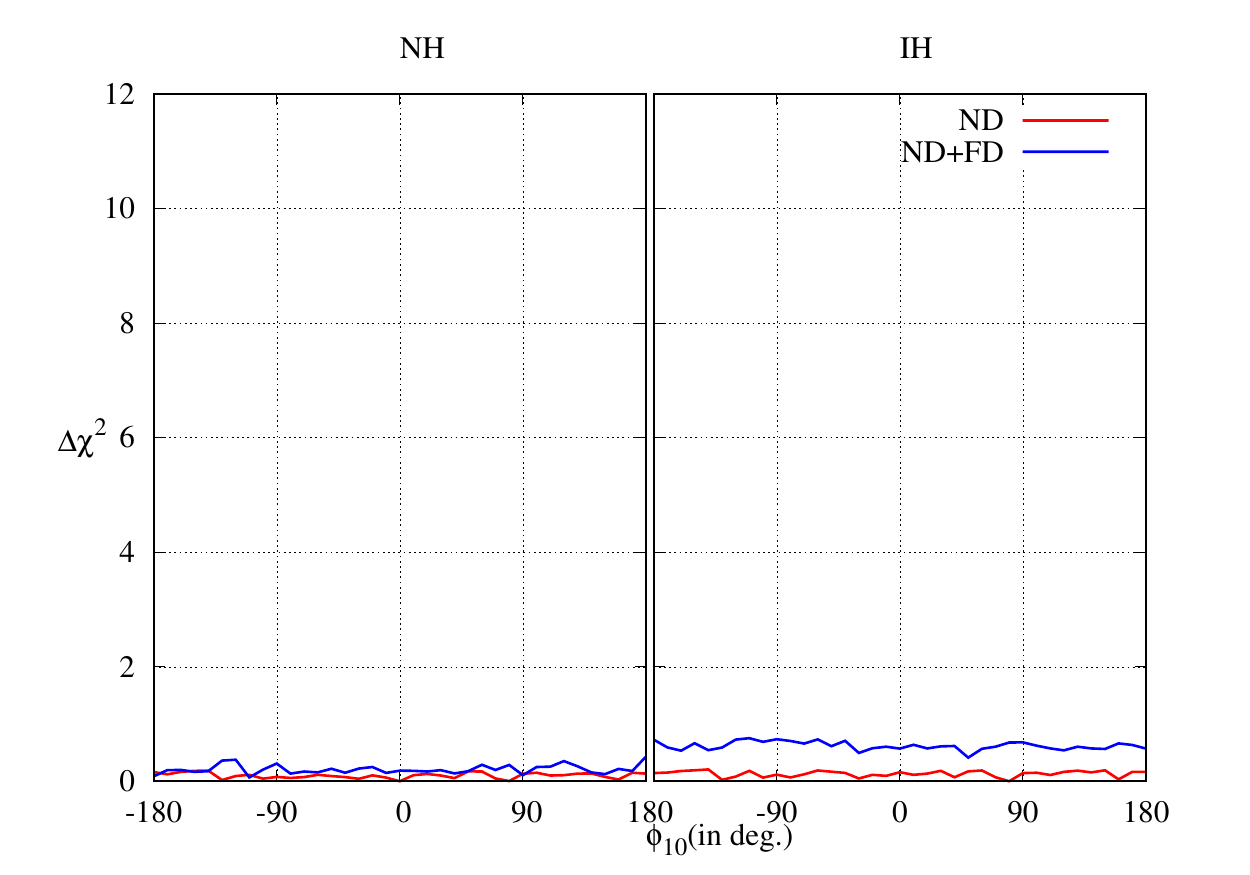}
\caption{\footnotesize{$\dchsq$ as a function of individual non-unitary parameters for the \nova ND (solid lines) and ND+FD (dashed lines) data.
}}
\label{alpha-precision}
\end{figure}

After fitting the \nova ND data with both the standard unitary and non-unitary mixing schemes, we found that the best-fit point occurs at $\alpha_{00}=0.998$, $|\alpha_{10}|=0.002$, and $\alpha_{11}=0.997$. In Fig.~\ref{alpha-precision} we show the $\dchsq$ values for the non-unitary parameters. The standard unitary oscillation gives as good fit as the non-unitary mixing to the data with $\dchsq=0.09$ and with $4$ degrees of freedom (d.o.f.) fewer. Therefore, it can be stated that the \nova ND data do not have any signature of non-unitary mixing. As mentioned earlier, the ND data from \nova do not have any sensitivity to the phase $\phi_{10}$, nor do they have any sensitivity to hierarchy. The $90\%$ C.L.\ ($5\,\sigma$) limit on the non-unitary parameters from the ND data of \nova are
\begin{equation}
    \alpha_{00}>0.911~(0.804),\, |\alpha_{10}|<0.020~(0.035),\, \alpha_{11}>0.952~(0.866)
\end{equation}
Fitting the combined data from the ND and FD, the best-fit points were found at $\alpha_{00}=0.991$, $|\alpha_{10}|=0.0009$, and $\alpha_{11}=0.978$. In this case also, the standard oscillation gives as good fit to the data as non-unitary mixing with $\dchsq=0.44$. Thus just like the ND data alone, the combined fit of the data from both ND and FD of \nova does not show any signature of non-unitary mixing (see Fig.~\ref{alpha-precision}). The $90\%$ C.L/\ ($5\,\sigma$) limit on the non-unitary parameters from the ND and FD data of \nova are
\begin{equation}
    \alpha_{00}>0.910~(0.804),\, |\alpha_{10}|<0.018~(0.035),\, \alpha_{11}>0.945~(0.885)
\end{equation}

\begin{figure}[htbp]
\vskip -1.5cm
\centering
\includegraphics[width=0.9\textwidth]{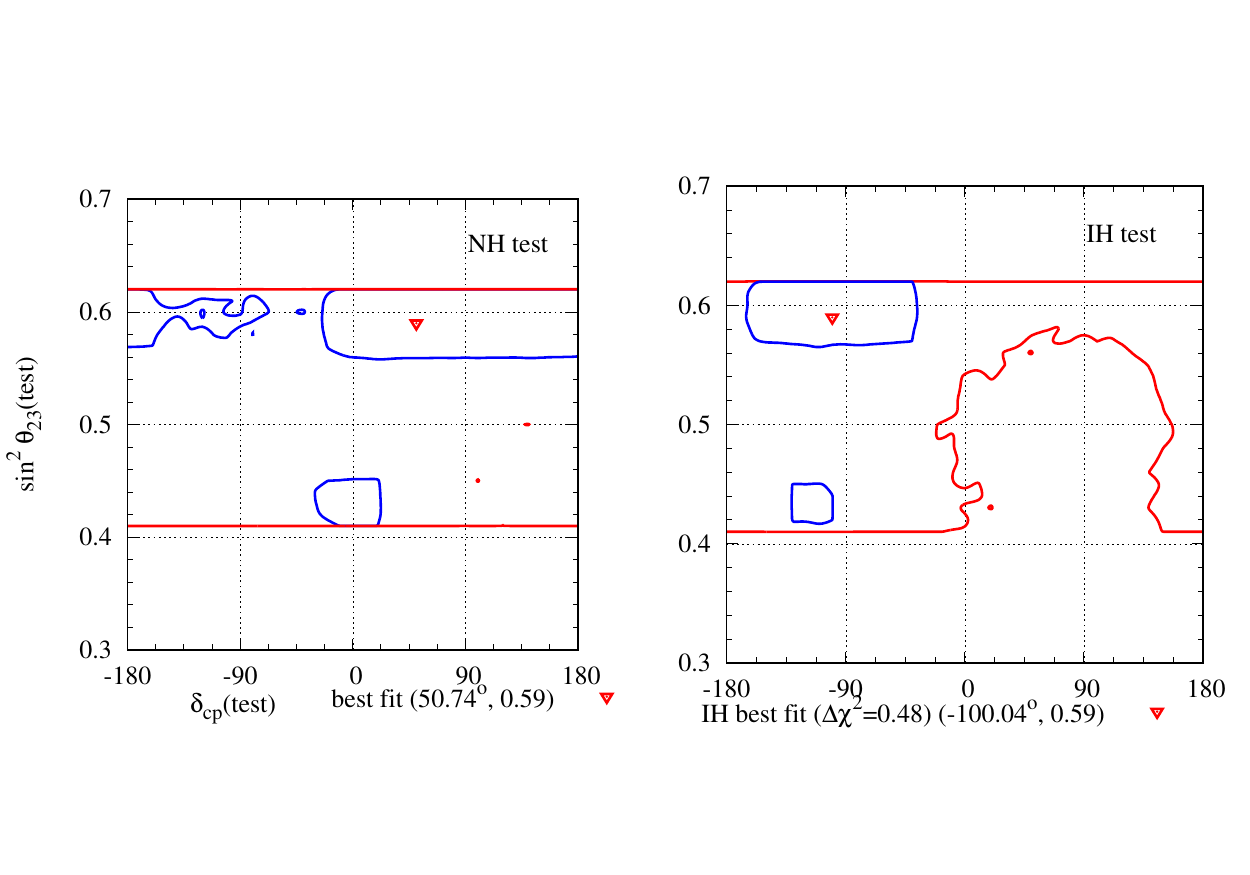}
\vskip -1.5cm
\caption{\footnotesize{Allowed regions in the $\sin^2 \tz-\dcp$ plane after combined analysis of the \nova ND and FD data. The blue (red) lines represent the boundary of $1\, \sigma$ ($3\, \sigma$) C.L. 
}}
\label{allowed}
\end{figure}

Although, as mentioned earlier, the ND data is not sensitive to the standard oscillation parameters, the combined ND and FD data is sensitive to those parameters. In Fig.~\ref{allowed}, we show the $1\,\sigma$ and $3\,\sigma$ allowed regions in the $\sin^2\tz-\dcp$ plane after the combined analysis. It is important to note that the best-fit points and the allowed regions are very close to each other for the non-unitary and standard unitary mixing. This is precisely because the sensitivity to the standard oscillation parameters is dominated by the FD data and the deviation from the unitarity is very small.
In Ref.~\cite{Miranda:2019ynh}, we did not include the \nova ND data, thus the deviation from non-unitarity at the FD data was large. Hence, the shift in the values of the standard unitary parameters with non-unitary mixing was large. It should also be noted in Fig.~\ref{allowed} that the IH cannot be ruled out even at $1\, \sigma$ C.L.\ and the hierarchy-$\dcp$ degeneracy is present in the data when analyzed with non-unitary mixing. 

\section{Conclusion}
\label{conclusion}
The \nova ND data does not show any signature of non-unitary mixing. At the best-fit point, the deviation from the unitary mixing is negligible. Although after combining the ND and FD data the deviation from the unitary mixing at the best-fit point is larger, the unitary mixing continues to give as good fit to the data as the non-unitary mixing. The limits obtained on the non-unitary parameters $\alpha_{00}$, $|\alpha_{10}|$, and $\alpha_{11}$ are  consistent with the limits obtained in Ref.~\cite{Forero:2021azc} from analyzing short baseline neutrino data.
Addition of the FD data gives sensitivity to the standard oscillation parameters. However, since the ND data allows only a small deviation from the unitary mixing, the results in the $\sin^2\tz-\dcp$ plane is almost similar to that obtained with the unitary mixing and with only the FD data. 

The results obtained from the ND data of \nova are in tension with the analysis of FD data alone as presented in Ref.~\cite{Miranda:2019ynh}. The FD data of \nova prefers non-unitary mixing over unitary mixing at $1\, \sigma$ C.L., and also the deviation from the unitarity at the best-fit point is quite large. Thus, it is important to fit the ND data of T2K as well as the future experiments like DUNE \cite{Abi:2018dnh} and T2HK \cite{Ishida:2013kba} to the non-unitary mixing scheme to investigate if similar trend persists in them as well.  

\section*{Acknowledgement}
We thank Alexander Himmel for valuable discussions and helps through private communications regarding the \nova near detector. This work was supported by a grant from the University of Johannesburg Research Council.

\bibliographystyle{apsrev}

\bibliography{referenceslist}

\begin{thebibliography}{27}
\expandafter\ifx\csname natexlab\endcsname\relax\def\natexlab#1{#1}\fi
\expandafter\ifx\csname bibnamefont\endcsname\relax
  \def\bibnamefont#1{#1}\fi
\expandafter\ifx\csname bibfnamefont\endcsname\relax
  \def\bibfnamefont#1{#1}\fi
\expandafter\ifx\csname citenamefont\endcsname\relax
  \def\citenamefont#1{#1}\fi
\expandafter\ifx\csname url\endcsname\relax
  \def\url#1{\texttt{#1}}\fi
\expandafter\ifx\csname urlprefix\endcsname\relax\def\urlprefix{URL }\fi
\providecommand{\bibinfo}[2]{#2}
\providecommand{\eprint}[2][]{\url{#2}}

\bibitem[{\citenamefont{Himmel}(2020)}]{Himmel:2020}
\bibinfo{author}{\bibfnamefont{A.}~\bibnamefont{Himmel}}
  (\bibinfo{year}{2020}), \bibinfo{note}{talk given at the Neutrino 2020
  meeting on July, 2nd, 2020,
  \url{https://indico.fnal.gov/event/43209/contributions/187840/attachments/130740/159597/NOvA-Oscilations-NEUTRINO2020.pdf}}.

\bibitem[{\citenamefont{Acero et~al.}(2021)}]{NOvA:2021nfi}
\bibinfo{author}{\bibfnamefont{M.~A.} \bibnamefont{Acero}} \bibnamefont{et~al.}
  (\bibinfo{collaboration}{NOvA}) (\bibinfo{year}{2021}), \eprint{2108.08219}.

\bibitem[{\citenamefont{Dunne}(2020)}]{Dunne:2020}
\bibinfo{author}{\bibfnamefont{P.}~\bibnamefont{Dunne}} (\bibinfo{year}{2020}),
  \bibinfo{note}{talk given at the Neutrino 2020 meeting on July, 2nd, 2020,
  \url{https://indico.fnal.gov/event/43209/contributions/187830/attachments/129636/159603/T2K_Neutrino2020.pdf}}.

\bibitem[{\citenamefont{Kelly et~al.}(2021)\citenamefont{Kelly, Machado, Parke,
  Perez-Gonzalez, and Funchal}}]{Kelly:2020fkv}
\bibinfo{author}{\bibfnamefont{K.~J.} \bibnamefont{Kelly}},
  \bibinfo{author}{\bibfnamefont{P.~A.~N.} \bibnamefont{Machado}},
  \bibinfo{author}{\bibfnamefont{S.~J.} \bibnamefont{Parke}},
  \bibinfo{author}{\bibfnamefont{Y.~F.} \bibnamefont{Perez-Gonzalez}},
  \bibnamefont{and} \bibinfo{author}{\bibfnamefont{R.~Z.}
  \bibnamefont{Funchal}}, \bibinfo{journal}{Phys. Rev. D}
  \textbf{\bibinfo{volume}{103}}, \bibinfo{pages}{013004}
  (\bibinfo{year}{2021}), \eprint{2007.08526}.

\bibitem[{\citenamefont{Chatterjee and Palazzo}(2021)}]{Chatterjee:2020kkm}
\bibinfo{author}{\bibfnamefont{S.~S.} \bibnamefont{Chatterjee}}
  \bibnamefont{and} \bibinfo{author}{\bibfnamefont{A.}~\bibnamefont{Palazzo}},
  \bibinfo{journal}{Phys. Rev. Lett.} \textbf{\bibinfo{volume}{126}},
  \bibinfo{pages}{051802} (\bibinfo{year}{2021}), \eprint{2008.04161}.

\bibitem[{\citenamefont{Denton et~al.}(2021)\citenamefont{Denton, Gehrlein, and
  Pestes}}]{Denton:2020uda}
\bibinfo{author}{\bibfnamefont{P.~B.} \bibnamefont{Denton}},
  \bibinfo{author}{\bibfnamefont{J.}~\bibnamefont{Gehrlein}}, \bibnamefont{and}
  \bibinfo{author}{\bibfnamefont{R.}~\bibnamefont{Pestes}},
  \bibinfo{journal}{Phys. Rev. Lett.} \textbf{\bibinfo{volume}{126}},
  \bibinfo{pages}{051801} (\bibinfo{year}{2021}), \eprint{2008.01110}.

\bibitem[{\citenamefont{Rahaman}(2021)}]{Rahaman:2021leu}
\bibinfo{author}{\bibfnamefont{U.}~\bibnamefont{Rahaman}}
  (\bibinfo{year}{2021}), \eprint{2103.04576}.

\bibitem[{\citenamefont{Miranda et~al.}(2021)\citenamefont{Miranda, Pasquini,
  Rahaman, and Razzaque}}]{Miranda:2019ynh}
\bibinfo{author}{\bibfnamefont{L.~S.} \bibnamefont{Miranda}},
  \bibinfo{author}{\bibfnamefont{P.}~\bibnamefont{Pasquini}},
  \bibinfo{author}{\bibfnamefont{U.}~\bibnamefont{Rahaman}}, \bibnamefont{and}
  \bibinfo{author}{\bibfnamefont{S.}~\bibnamefont{Razzaque}},
  \bibinfo{journal}{Eur. Phys. J. C} \textbf{\bibinfo{volume}{81}},
  \bibinfo{pages}{444} (\bibinfo{year}{2021}), \eprint{1911.09398}.

\bibitem[{\citenamefont{Aguilar-Arevalo et~al.}(2001)}]{Aguilar:2001ty}
\bibinfo{author}{\bibfnamefont{A.}~\bibnamefont{Aguilar-Arevalo}}
  \bibnamefont{et~al.} (\bibinfo{collaboration}{LSND}), \bibinfo{journal}{Phys.
  Rev.} \textbf{\bibinfo{volume}{D64}}, \bibinfo{pages}{112007}
  (\bibinfo{year}{2001}), \eprint{hep-ex/0104049}.

\bibitem[{\citenamefont{Aguilar-Arevalo
  et~al.}(2018)}]{Aguilar-Arevalo:2018gpe}
\bibinfo{author}{\bibfnamefont{A.~A.} \bibnamefont{Aguilar-Arevalo}}
  \bibnamefont{et~al.} (\bibinfo{collaboration}{MiniBooNE}),
  \bibinfo{journal}{Phys. Rev. Lett.} \textbf{\bibinfo{volume}{121}},
  \bibinfo{pages}{221801} (\bibinfo{year}{2018}), \eprint{1805.12028}.

\bibitem[{\citenamefont{Forero et~al.}(2021)\citenamefont{Forero, Giunti,
  Ternes, and Tortola}}]{Forero:2021azc}
\bibinfo{author}{\bibfnamefont{D.~V.} \bibnamefont{Forero}},
  \bibinfo{author}{\bibfnamefont{C.}~\bibnamefont{Giunti}},
  \bibinfo{author}{\bibfnamefont{C.~A.} \bibnamefont{Ternes}},
  \bibnamefont{and} \bibinfo{author}{\bibfnamefont{M.}~\bibnamefont{Tortola}}
  (\bibinfo{year}{2021}), \eprint{2103.01998}.

\bibitem[{\citenamefont{Escrihuela et~al.}(2015)\citenamefont{Escrihuela,
  Forero, Miranda, Tortola, and Valle}}]{Escrihuela:2015wra}
\bibinfo{author}{\bibfnamefont{F.~J.} \bibnamefont{Escrihuela}},
  \bibinfo{author}{\bibfnamefont{D.~V.} \bibnamefont{Forero}},
  \bibinfo{author}{\bibfnamefont{O.~G.} \bibnamefont{Miranda}},
  \bibinfo{author}{\bibfnamefont{M.}~\bibnamefont{Tortola}}, \bibnamefont{and}
  \bibinfo{author}{\bibfnamefont{J.~W.~F.} \bibnamefont{Valle}},
  \bibinfo{journal}{Phys. Rev.} \textbf{\bibinfo{volume}{D92}},
  \bibinfo{pages}{053009} (\bibinfo{year}{2015}), \bibinfo{note}{[Erratum:
  Phys. Rev.D93,no.11,119905(2016)]}, \eprint{1503.08879}.

\bibitem[{\citenamefont{Astier et~al.}(2003)}]{NOMAD:2003mqg}
\bibinfo{author}{\bibfnamefont{P.}~\bibnamefont{Astier}} \bibnamefont{et~al.}
  (\bibinfo{collaboration}{NOMAD}), \bibinfo{journal}{Phys. Lett. B}
  \textbf{\bibinfo{volume}{570}}, \bibinfo{pages}{19} (\bibinfo{year}{2003}),
  \eprint{hep-ex/0306037}.

\bibitem[{\citenamefont{Avvakumov et~al.}(2002)}]{NuTeV:2002daf}
\bibinfo{author}{\bibfnamefont{S.}~\bibnamefont{Avvakumov}}
  \bibnamefont{et~al.} (\bibinfo{collaboration}{NuTeV}),
  \bibinfo{journal}{Phys. Rev. Lett.} \textbf{\bibinfo{volume}{89}},
  \bibinfo{pages}{011804} (\bibinfo{year}{2002}), \eprint{hep-ex/0203018}.

\bibitem[{\citenamefont{Dentler et~al.}(2017)\citenamefont{Dentler,
  Hern\'andez-Cabezudo, Kopp, Maltoni, and Schwetz}}]{Dentler:2017tkw}
\bibinfo{author}{\bibfnamefont{M.}~\bibnamefont{Dentler}},
  \bibinfo{author}{\bibfnamefont{A.}~\bibnamefont{Hern\'andez-Cabezudo}},
  \bibinfo{author}{\bibfnamefont{J.}~\bibnamefont{Kopp}},
  \bibinfo{author}{\bibfnamefont{M.}~\bibnamefont{Maltoni}}, \bibnamefont{and}
  \bibinfo{author}{\bibfnamefont{T.}~\bibnamefont{Schwetz}},
  \bibinfo{journal}{JHEP} \textbf{\bibinfo{volume}{11}}, \bibinfo{pages}{099}
  (\bibinfo{year}{2017}), \eprint{1709.04294}.

\bibitem[{\citenamefont{{P. Dunne}}({2021})}]{Dunne-com}
\bibinfo{author}{\bibnamefont{{P. Dunne}}}, \bibinfo{howpublished}{{private
  communication}} (\bibinfo{year}{{2021}}).

\bibitem[{\citenamefont{Acero et~al.}(2018)}]{NOvA:2018gge}
\bibinfo{author}{\bibfnamefont{M.~A.} \bibnamefont{Acero}} \bibnamefont{et~al.}
  (\bibinfo{collaboration}{NOvA}), \bibinfo{journal}{Phys. Rev. D}
  \textbf{\bibinfo{volume}{98}}, \bibinfo{pages}{032012}
  (\bibinfo{year}{2018}), \eprint{1806.00096}.

\bibitem[{\citenamefont{Antusch and Fischer}(2014)}]{Antusch_2014}
\bibinfo{author}{\bibfnamefont{S.}~\bibnamefont{Antusch}} \bibnamefont{and}
  \bibinfo{author}{\bibfnamefont{O.}~\bibnamefont{Fischer}},
  \bibinfo{journal}{Journal of High Energy Physics}
  \textbf{\bibinfo{volume}{2014}} (\bibinfo{year}{2014}), ISSN
  \bibinfo{issn}{1029-8479},
  \urlprefix\url{http://dx.doi.org/10.1007/JHEP10(2014)094}.

\bibitem[{\citenamefont{Escrihuela et~al.}(2017)\citenamefont{Escrihuela,
  Forero, Miranda, Tórtola, and Valle}}]{Escrihuela:2016ube}
\bibinfo{author}{\bibfnamefont{F.~J.} \bibnamefont{Escrihuela}},
  \bibinfo{author}{\bibfnamefont{D.~V.} \bibnamefont{Forero}},
  \bibinfo{author}{\bibfnamefont{O.~G.} \bibnamefont{Miranda}},
  \bibinfo{author}{\bibfnamefont{M.}~\bibnamefont{Tórtola}}, \bibnamefont{and}
  \bibinfo{author}{\bibfnamefont{J.~W.~F.} \bibnamefont{Valle}},
  \bibinfo{journal}{New J. Phys.} \textbf{\bibinfo{volume}{19}},
  \bibinfo{pages}{093005} (\bibinfo{year}{2017}), \eprint{1612.07377}.

\bibitem[{\citenamefont{Huber et~al.}(2005)\citenamefont{Huber, Lindner, and
  Winter}}]{Huber:2004ka}
\bibinfo{author}{\bibfnamefont{P.}~\bibnamefont{Huber}},
  \bibinfo{author}{\bibfnamefont{M.}~\bibnamefont{Lindner}}, \bibnamefont{and}
  \bibinfo{author}{\bibfnamefont{W.}~\bibnamefont{Winter}},
  \bibinfo{journal}{Comput.Phys.Commun.} \textbf{\bibinfo{volume}{167}},
  \bibinfo{pages}{195} (\bibinfo{year}{2005}), \eprint{hep-ph/0407333}.

\bibitem[{\citenamefont{Huber et~al.}(2007)\citenamefont{Huber, Kopp, Lindner,
  Rolinec, and Winter}}]{Huber:2007ji}
\bibinfo{author}{\bibfnamefont{P.}~\bibnamefont{Huber}},
  \bibinfo{author}{\bibfnamefont{J.}~\bibnamefont{Kopp}},
  \bibinfo{author}{\bibfnamefont{M.}~\bibnamefont{Lindner}},
  \bibinfo{author}{\bibfnamefont{M.}~\bibnamefont{Rolinec}}, \bibnamefont{and}
  \bibinfo{author}{\bibfnamefont{W.}~\bibnamefont{Winter}},
  \bibinfo{journal}{Comput.Phys.Commun.} \textbf{\bibinfo{volume}{177}},
  \bibinfo{pages}{432} (\bibinfo{year}{2007}), \eprint{hep-ph/0701187}.

\bibitem[{\citenamefont{Esteban et~al.}(2019)\citenamefont{Esteban,
  Gonzalez-Garcia, Hernandez-Cabezudo, Maltoni, and Schwetz}}]{Esteban:2018azc}
\bibinfo{author}{\bibfnamefont{I.}~\bibnamefont{Esteban}},
  \bibinfo{author}{\bibfnamefont{M.~C.} \bibnamefont{Gonzalez-Garcia}},
  \bibinfo{author}{\bibfnamefont{A.}~\bibnamefont{Hernandez-Cabezudo}},
  \bibinfo{author}{\bibfnamefont{M.}~\bibnamefont{Maltoni}}, \bibnamefont{and}
  \bibinfo{author}{\bibfnamefont{T.}~\bibnamefont{Schwetz}},
  \bibinfo{journal}{JHEP} \textbf{\bibinfo{volume}{01}}, \bibinfo{pages}{106}
  (\bibinfo{year}{2019}), \eprint{1811.05487}.

\bibitem[{\citenamefont{Nichol}(2012)}]{Nichol:2012}
\bibinfo{author}{\bibfnamefont{R.}~\bibnamefont{Nichol}}
  (\bibinfo{collaboration}{MINOS}) (\bibinfo{year}{2012}), \bibinfo{note}{talk
  given at the Neutrino 2012 Conference, June 3-9, 2012, Kyoto, Japan,
  \url{http://neu2012.kek.jp/}}.

\bibitem[{\citenamefont{Nunokawa et~al.}(2005)\citenamefont{Nunokawa, Parke,
  and Zukanovich~Funchal}}]{Nunokawa:2005nx}
\bibinfo{author}{\bibfnamefont{H.}~\bibnamefont{Nunokawa}},
  \bibinfo{author}{\bibfnamefont{S.~J.} \bibnamefont{Parke}}, \bibnamefont{and}
  \bibinfo{author}{\bibfnamefont{R.}~\bibnamefont{Zukanovich~Funchal}},
  \bibinfo{journal}{Phys.Rev.} \textbf{\bibinfo{volume}{D72}},
  \bibinfo{pages}{013009} (\bibinfo{year}{2005}), \eprint{hep-ph/0503283}.

\bibitem[{\citenamefont{Nufit}(2019)}]{nufit}
\bibinfo{author}{\bibnamefont{Nufit}} (\bibinfo{year}{2019}),
  \bibinfo{note}{\url{http://www.nu-fit.org/?q=node/211}}.

\bibitem[{\citenamefont{Abi et~al.}(2018)}]{Abi:2018dnh}
\bibinfo{author}{\bibfnamefont{B.}~\bibnamefont{Abi}} \bibnamefont{et~al.}
  (\bibinfo{collaboration}{DUNE}) (\bibinfo{year}{2018}), \eprint{1807.10334}.

\bibitem[{\citenamefont{Ishida}(2013)}]{Ishida:2013kba}
\bibinfo{author}{\bibfnamefont{T.}~\bibnamefont{Ishida}}
  (\bibinfo{collaboration}{Hyper-Kamiokande Working Group}), in
  \emph{\bibinfo{booktitle}{{15th International Workshop on Neutrino Factories,
  Super Beams and Beta Beams}}} (\bibinfo{year}{2013}), \eprint{1311.5287}.

\end{thebibliography}

\appendix*
\section{Oscillation Probabilities and fits to event distributions}

\begin{figure}[htbp]
\centering
\includegraphics[width=0.8\textwidth]{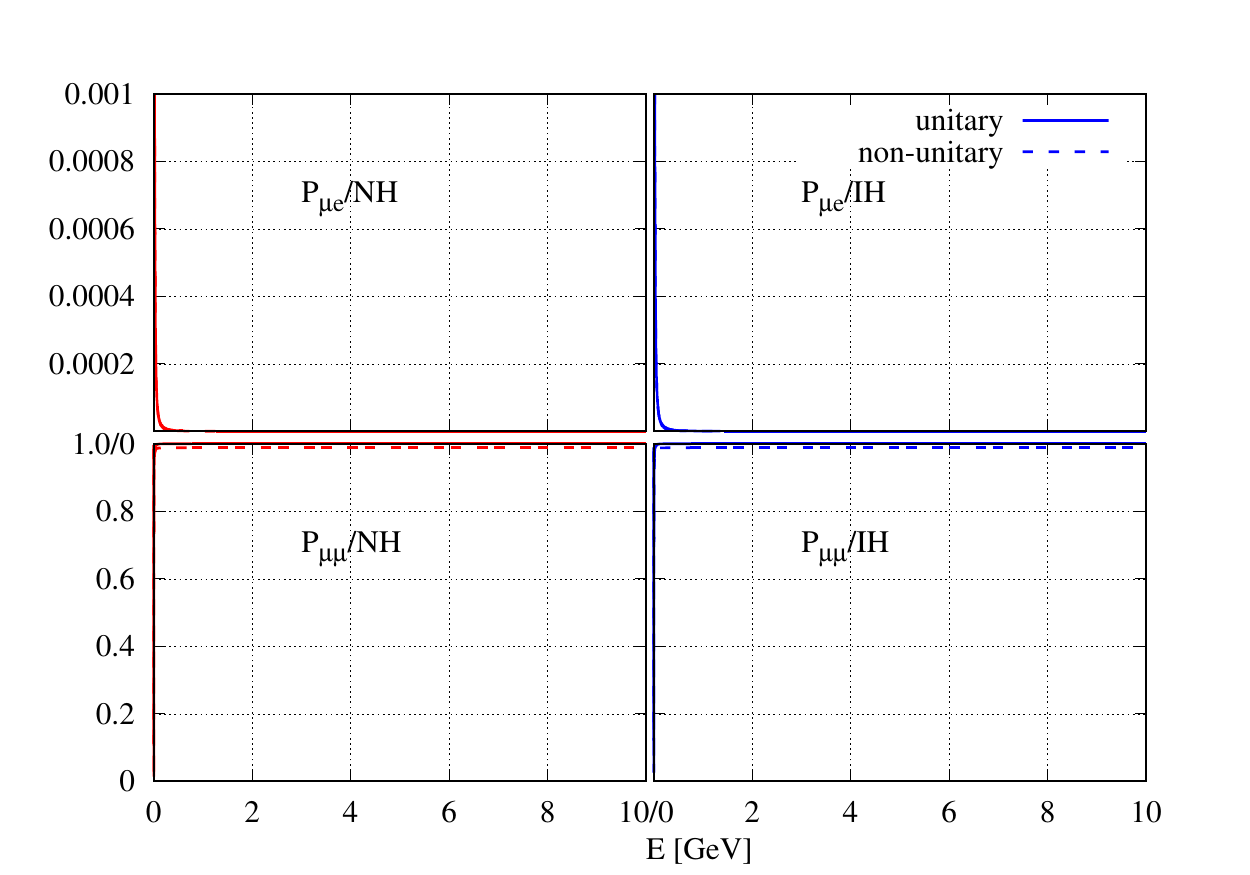}
\caption{\footnotesize{$\nu_\mu \to \nu_e$ transition ($\nu_\mu \to \nu_\mu$ survival) probability as a function of energy at the ND in the upper (lower) panel. The left (right) panel is for NH (IH). The solid (dashed) line represents (non-) unitary mixing.
}}
\label{prob-nu-ND}
\end{figure}

\begin{figure}[htbp]
\centering
\includegraphics[width=0.8\textwidth]{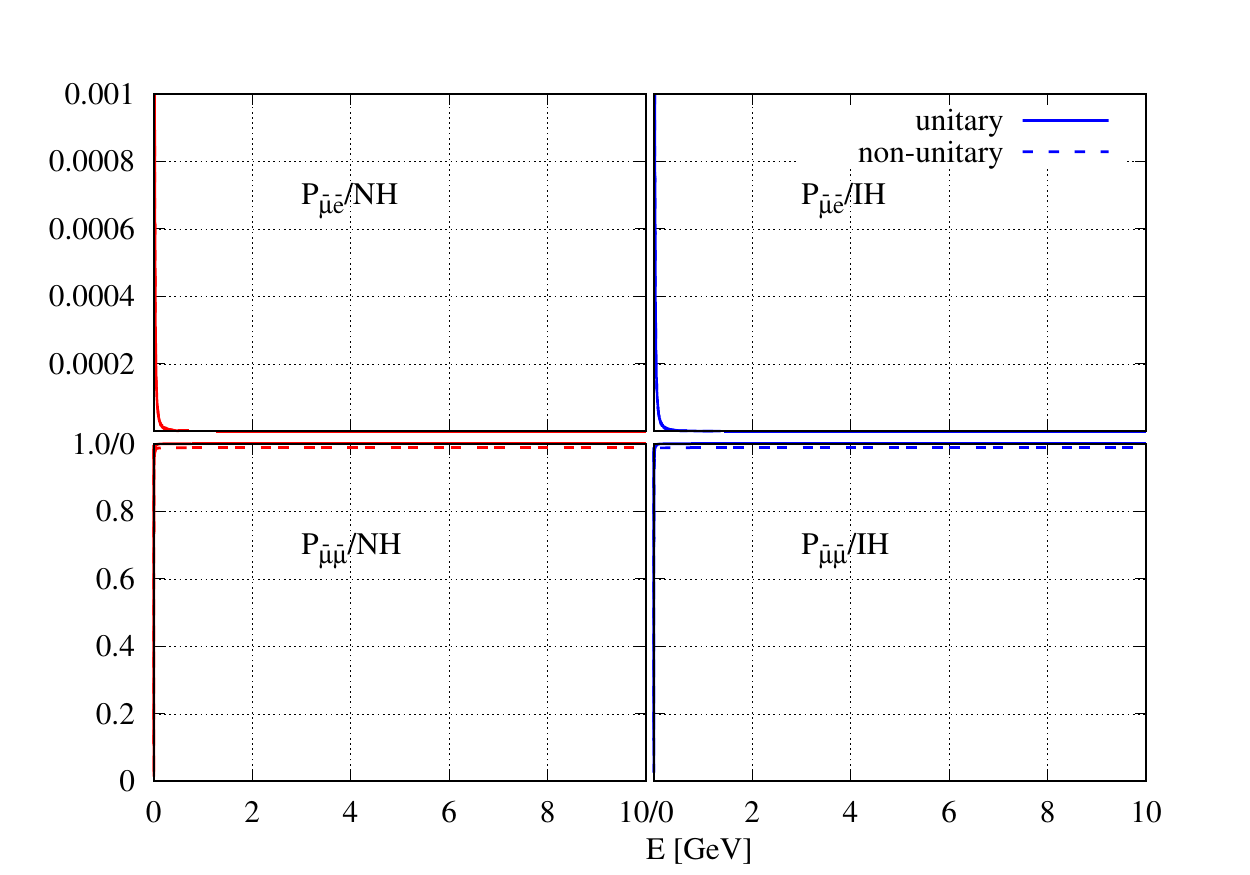}
\caption{\footnotesize{
The same as Fig.~\ref{prob-nu-ND} but for antineutrinos.
}}
\label{prob-nubar-ND}
\end{figure}

In Fig.~\ref{prob-nu-ND}, we have presented the $\nu_\mu \to \nu_e$ transition probability and $\nu_\mu \to \nu_\mu$ survival probability as functions of energy at the ND for both the unitary and non-unitary mixing schemes. To do so, we changed the probability code in GLoBES such that it can calculate the exact $\nu_e$ appearance and $\nu_\mu$ survival probabilities with non-unitary mixing and without any approximation at a particular baseline, however small. This is discussed in details in Ref.~\cite{Miranda:2019ynh} We fixed the standard unitary oscillation parameter values at their current \nova best-fit values. The non-unitary parameter values have been fixed at the best-fit values we got after analyzing the ND data from NO$\nu$A. It is obvious from the figure that for unitary mixing at the ND, the $\nu_\mu \to \nu_e$ appearance probability is $0$, and $\nu_\mu \to \nu_\mu$ survival probability is 1 for all the energy. At the present best-fit point found after fitting the ND data, the oscillation and survival probabilities do not change. The characteristics do not change for $\bar{\nu}$ probability shown in Fig.~\ref{prob-nubar-ND}.

\begin{figure}[htbp]
\centering
\includegraphics[width=0.8\textwidth]{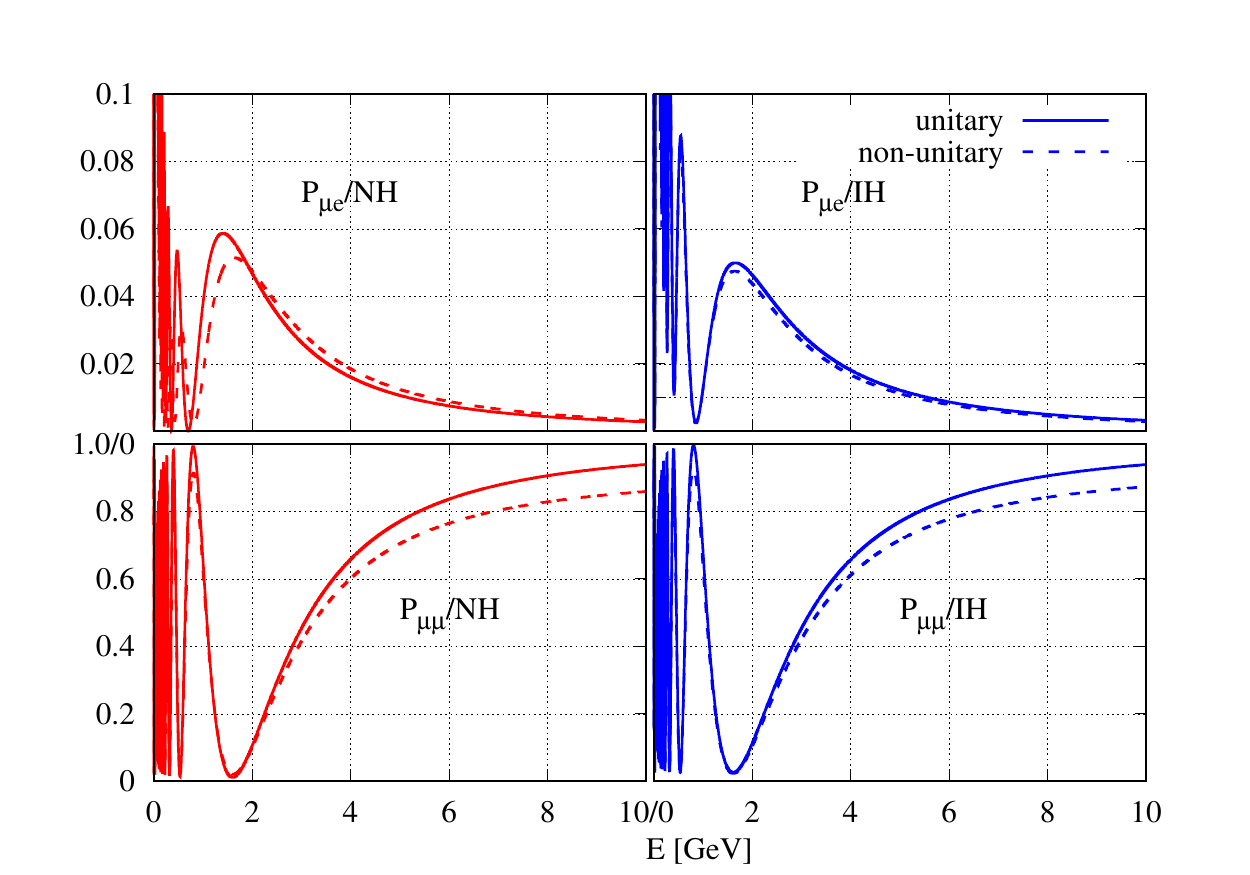}
\caption{\footnotesize{
$\nu_\mu \to \nu_e$ transition ($\nu_\mu \to \nu_\mu$ survival) probability as a function of energy at the FD in the upper (lower) panel. The left (right) panel is for NH (IH). The solid (dashed) line represents (non-) unitary mixing.
}}
\label{prob-nu-FD}
\end{figure}

\begin{figure}[htbp]
\centering
\includegraphics[width=0.8\textwidth]{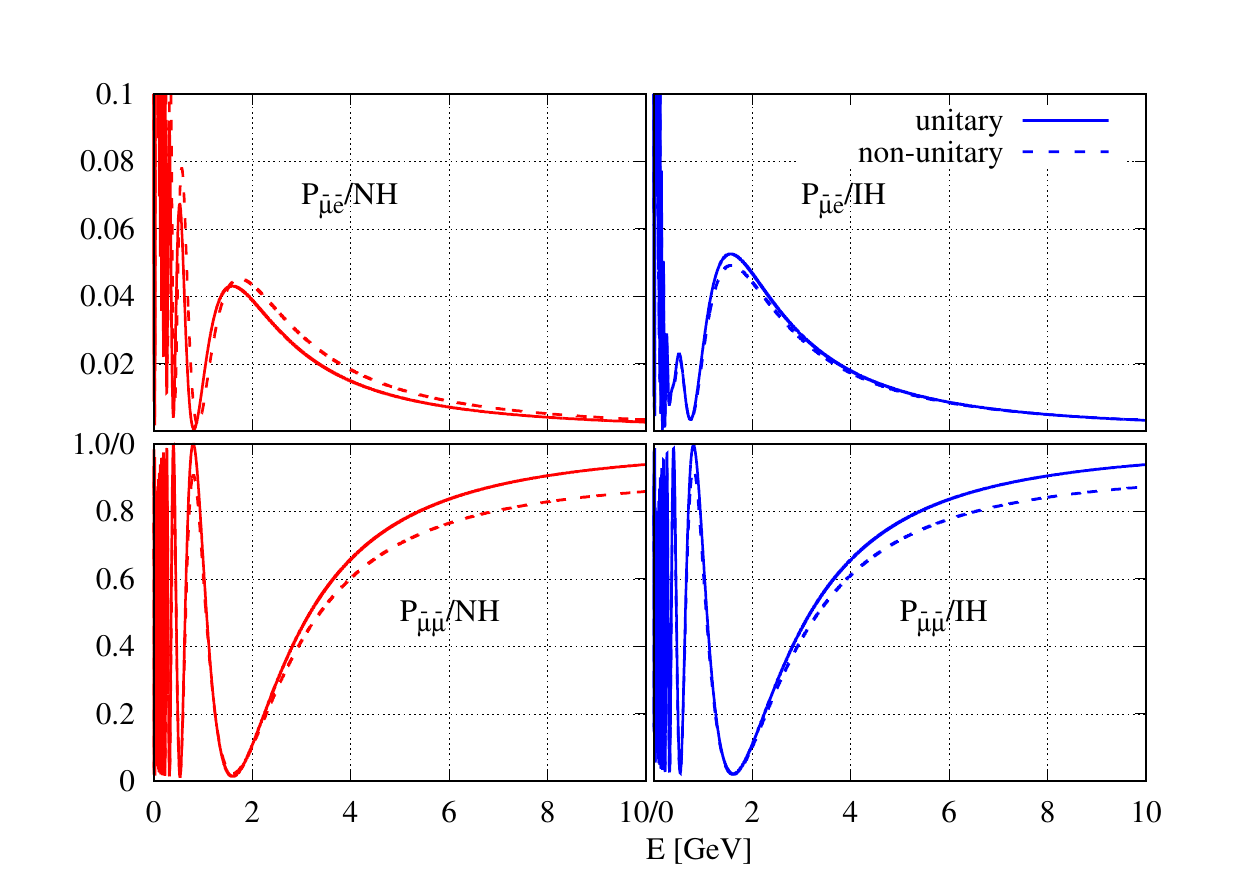}
\caption{\footnotesize{
The same as Fig.~\ref{prob-nu-FD} but for antineutrinos.
}}
\label{prob-nubar-FD}
\end{figure}

In Figs.~\ref{prob-nu-FD} and \ref{prob-nubar-FD}, we have shown the same probabilities at the FD. For the unitary mixing, we have fixed the standard oscillation probabilities at the present best-fit values given by the \nova Collaboration. For non-unitary mixing, we have fixed both the standard and non-standard oscillation parameters at the best-fit values we have obtained after analyzing the ND and FD data with non-unitary mixing hypothesis. It is obvious that the distinction between the two hypotheses at the FD is small at the present best-fit values.

\begin{figure}[htbp]
\centering
\includegraphics[width=0.8\textwidth]{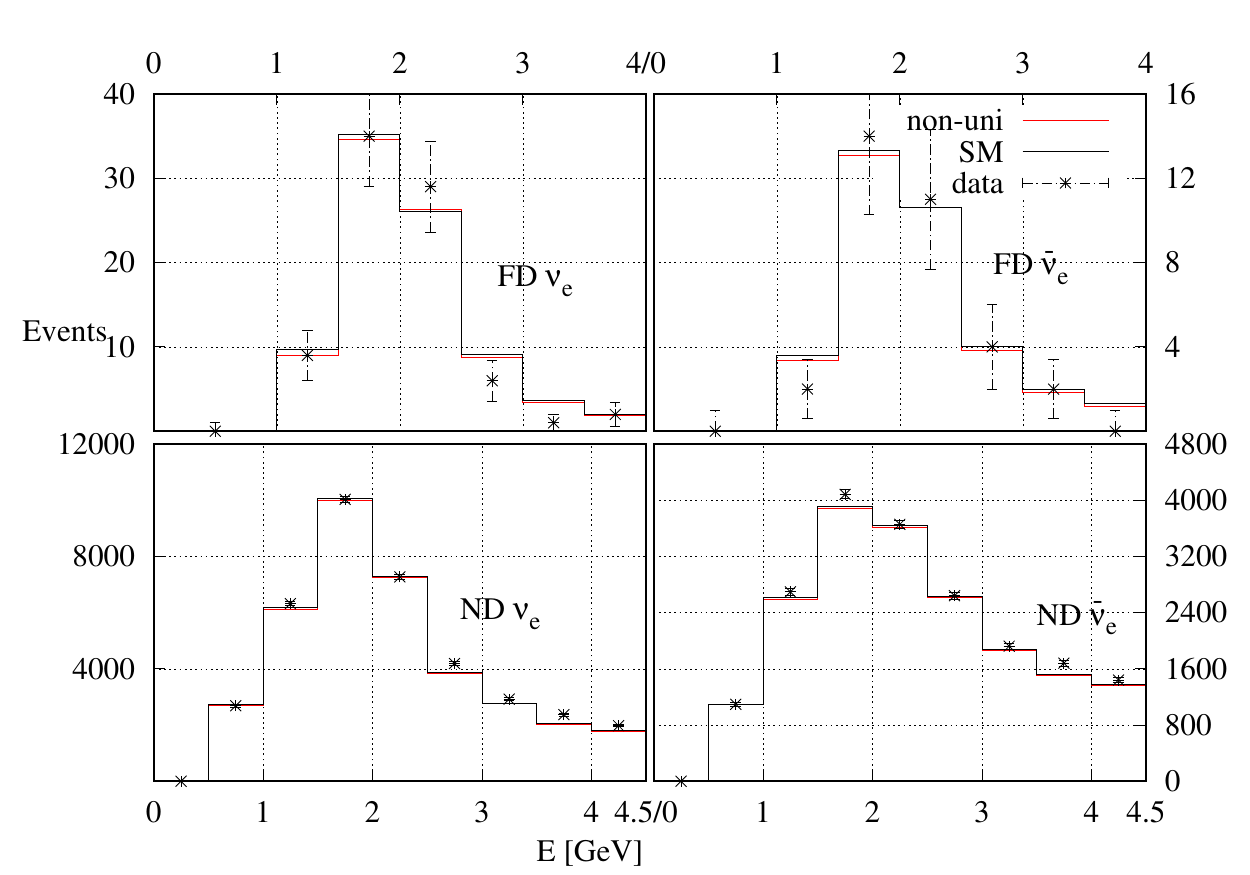}
\caption{\footnotesize{Comparison between the observed and expected $\nu_e$ and $\bar{\nu}_e$ events at their respective best-fit points for both unitary and non-unitary mixing at the ND and FD.
}}
\label{events}
\end{figure}

\begin{figure}[htbp]
\centering
\includegraphics[width=0.8\textwidth]{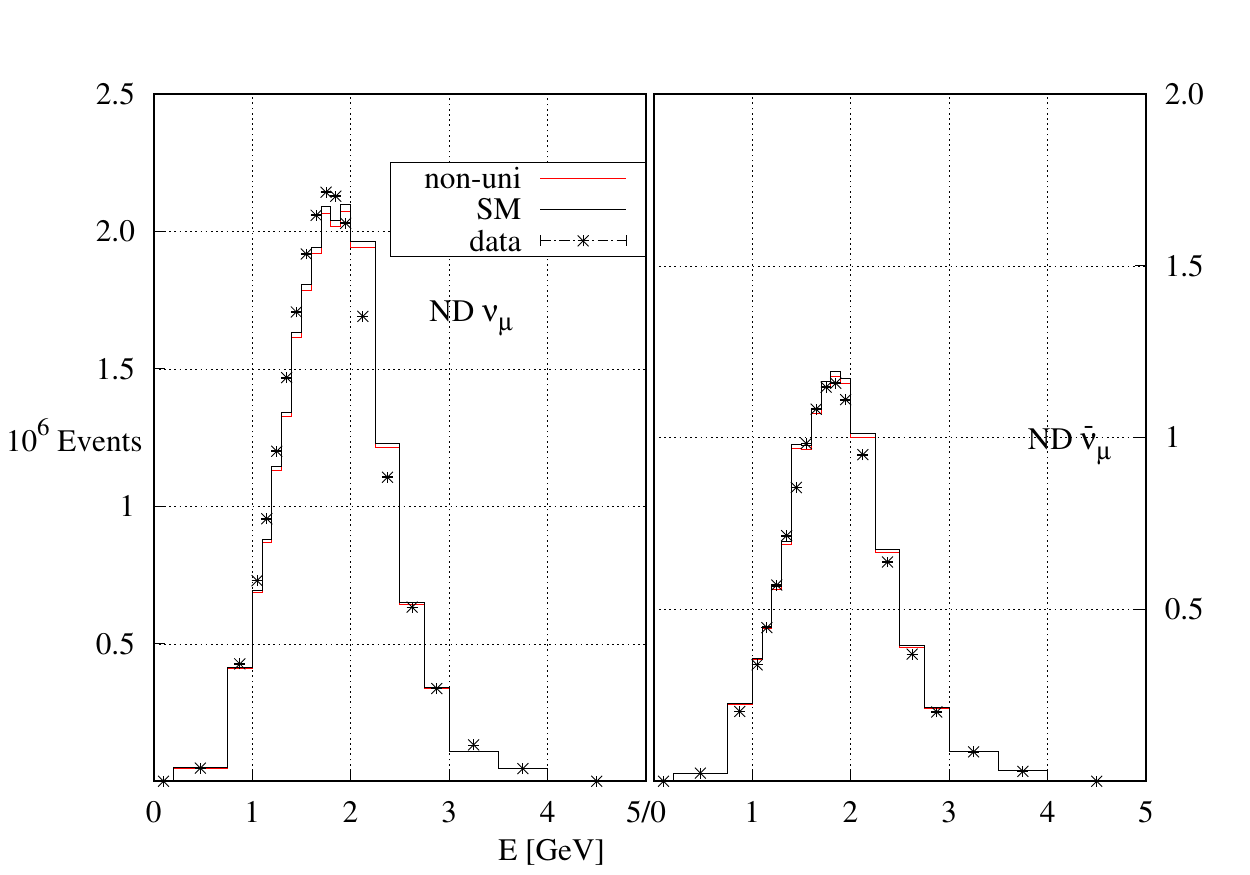}
\caption{\footnotesize{
Same as Fig.~\ref{events} 
but for $\nu_\mu$ and $\bar{\nu}_\mu$ events.
}}
\label{events-disapp-ND}
\end{figure}

In Fig.~\ref{events} we have shown the comparison between the observed $\nu_e$ and $\bar{\nu}_e$ events at the \nova ND and FD with the expected events for the standard unitary and non-unitary mixing schemes. Expected events are calculated with mixing parameters at the respective best-fit points of the two schemes. The difference in expected event numbers at the FD between unitary and non-unitary mixing is very small. At ND, the expected event numbers at the best-fit points for both the mixing hypotheses are essentially same and both of them give a good fit to the data. In Fig.~\ref{events-disapp-ND}, we have shown the difference between the expected and observed $\nu_\mu$ and $\bar{\nu}_\mu$ events at the ND as a function of energy for both the unitary and non-unitary mixing.

\end{document}